# Magnetic and Structural Properties of A-Site Ordered Chromium Spinel Sulfides: Alternating Antiferromagnetic and Ferromagnetic Interactions in the Breathing Pyrochlore Lattice

Yoshihiko Okamoto[1,*], Masaki Mori[1], Naoyuki Katayama[1], Atsushi Miyake[2], Masashi Tokunaga[2],
Akira Matsuo[2], Koichi Kindo[2], and Koshi Takenaka[1]

[1]*Department of Applied Physics, Nagoya University, Nagoya 464-8603, Japan*
[2]*Institute for Solid State Physics, University of Tokyo, Kashiwa 277-8581, Japan*

We report a comprehensive study on the magnetic and structural properties of the spinel sulfides $LiInCr_4S_8$, $LiGaCr_4S_8$, and $CuInCr_4S_8$, where $Li^+/Cu^+$ and $Ga^{3+}/In^{3+}$ ions form a zinc-blende-type order. On the basis of synchrotron X-ray diffraction and magnetization data obtained using polycrystalline samples, these three sulfides are suggested to be breathing pyrochlore magnets with alternating antiferromagnetic and ferromagnetic interactions on the small and large tetrahedra, respectively. The measured magnetization processes of the three sulfides up to 72 T are significantly different. The magnetization curves of $LiInCr_4S_8$ and $CuInCr_4S_8$ have large hysteresis loops with different shapes, while there is no hysteresis in that of $LiGaCr_4S_8$. Geometrical frustration of the small tetrahedron is likely to give rise to a wide variety of ground states, indicating the rich physics in these antiferromagnetic-ferromagnetic breathing pyrochlore magnets.

## I. INTRODUCTION

Breathing pyrochlore magnets, where localized spins are arranged at the edge of alternating small and large tetrahedra with magnetic interactions $J$ and $J'$, respectively, have attracted attention as unique systems that possess both geometrical frustration and bond alternation. The two systems that are spin-3/2 magnets, $LiInCr_4O_8$ and $LiGaCr_4O_8$ [1–9], and a pseudospin-1/2 magnet, $Ba_3Yb_2Zn_5O_{11}$ [10–12], have been intensively studied. Both systems crystallize with cubic $F$–$43m$ symmetry. $LiInCr_4O_8$ and $LiGaCr_4O_8$ are chromium spinel oxides, where $Li^+$ and $Ga^{3+}/In^{3+}$ ions form a zinc-blende-type order [1,2]. They are antiferromagnets with very large, negative Weiss temperatures of −332 and −659 K, respectively, suggesting that both $J$ and $J'$ are antiferromagnetic ($J, J' > 0$) [2]. The ratio of $J$ and $J'$ is considerably different between $LiInCr_4O_8$ and $LiGaCr_4O_8$. Their $J'/J$ values are estimated to be ~0.1 and ~0.6, respectively, as derived using an empirical relationship between the strength of the magnetic interactions and the Cr-Cr distances [2]. $LiInCr_4O_8$ shows spin-gap behavior caused by spin-singlet formation in the small tetrahedra above ~15 K, reflecting the small $J'/J$, while $LiGaCr_4O_8$ shows an antiferromagnetic short-range order below ~45 K in magnetic susceptibility data, much like conventional Cr spinel oxides such as $ZnCr_2O_4$. Both compounds also exhibit an antiferromagnetic long-range order at ~15 K, accompanied by a structural transition [3–5]. Various intriguing magnetic phenomena, such as the sup-pression of magnetic order and a transition to the possible spin-nematic phase, are induced by a slight chemical substitution to $LiInCr_4O_8$ and $LiGaCr_4O_8$, respectively [6,7]. Alternatively, $Ba_3Yb_2Zn_5O_{11}$ with $J \gg J'$ does not show any long-range magnetic order down to the lowest measured temperature of 0.38 K, but instead goes to the singlet ground state at low temperatures [10,11]. Since this singlet state is doubly degenerate due to the $T_d$ symmetry of a regular tetrahedron, it is interesting how this degeneracy is lifted when a finite $J'$ is introduced [13,14].

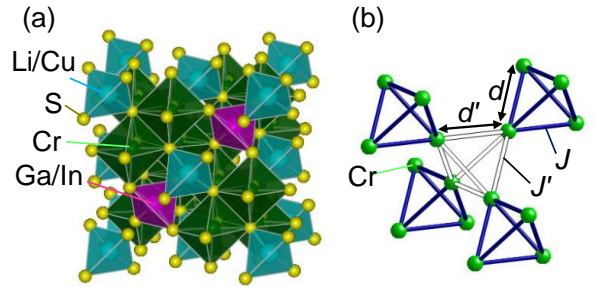

Fig. 1. (a) Crystal structure of $LiInCr_4S_8$, $LiGaCr_4S_8$, and $CuInCr_4S_8$, and (b) breathing pyrochlore lattice made of $Cr^{3+}$ ions in these three sulfides.

Here we focus on breathing pyrochlore magnets with antiferromagnetic $J$ and ferromagnetic $J'$, different from the aforementioned breathing pyrochlore magnets that have antiferromagnetic $J$ and $J'$. Benton and Shannon noted that the ground state of the former magnets is a collinear order with decoupled (001) ferromagnetic planes for the classical Heisenberg spin case [15]. It is still unclear, however, what kind of magnetic state appears in the real materials, compared with the latter magnets. In this paper, we report a comprehensive study on the magnetic and structural properties of the chromium spinel sulfides $LiInCr_4S_8$, $LiGaCr_4S_8$, and $CuInCr_4S_8$, shown in Fig. 1. These spinel sulfides were first synthesized by Pinch *et al*. in 1970 [16]. They reported that their magnetizations show anomalies at low temperatures, expected to correspond to a magnetic order, and the $Li^+/Cu^+$ and $Ga^{3+}/In^{3+}$ ions occupying tetrahedral sites form a zinc-blend-type order, as in the case of the oxides. $CuInCr_4S_8$ was intensively studied forty years ago as a related material of a room-temperature ferromagnet $CuCr_2S_4$ and reported to show a negative Weiss temperature of −77 or −100 K [17,18]. Magnetic Bragg peaks, indicating the presence of an antiferromagnetic long-range order, are observed in the neutron diffraction data measured at 4.2 K [17,19]. The proposed spin-structure model of the ordered phase is consistent with that of the collinear order mentioned above. A metamagnetic hysteresis appears above 25 T in high-field magnetization curves measured up to 38 T [20]. In contrast to $CuInCr_4S_8$, there has been no detailed study on $LiInCr_4S_8$ and $LiGaCr_4S_8$ thus far.

We find that these three sulfides are breathing pyrochlore magnets with antiferromagnetic $J$ and ferromagnetic $J'$ using the structural and magnetic parameters obtained by structural analyses using synchrotron powder X-ray diffraction (XRD) data and a Curie-Weiss fit to the magnetic susceptibility data, respectively. However, the low-temperature magnetic properties of these three sulfides differ significantly. Geometrical frustration of the small tetrahedra with antiferromagnetic $J$ is suggested to have a considerable effect on the magnetic properties of the antiferromagnetic-ferromagnetic breathing pyrochlore spin system, indicating that this system has rich physics comparable to that of an antiferromagnetic one.

## II. EXPERIMENTAL METHODS

Polycrystalline samples of $LiInCr_4S_8$, $LiGaCr_4S_8$, and $CuInCr_4S_8$ were prepared by a solid-state reaction method. A stoichiometric mixture of $Li_2S/Cu$, $Ga_2S_3/In_2S_3$, Cr, and S powders was sealed in an evacuated quartz tube. $Li_2S$ was handled in an inert gas atmosphere, because it is moisture sensitive. The tube was first kept at 673 K for 24 h to prevent the fast sublimation of sulfur, and then 1023 K for 96 h for $LiInCr_4S_8$, 1023 K for 48 h and then 1323 K for 48 h for $LiGaCr_4S_8$, and 1023 K for 144 h for $CuInCr_4S_8$ with intermediate grindings.

The crystal structures of $LiInCr_4S_8$, $LiGaCr_4S_8$, and $CuInCr_4S_8$ were determined by Rietveld analyses for the powder XRD patterns obtained at room temperature by employing synchrotron X-ray with λ = 0.652256 Å at BL5S2 at the Aichi Synchrotron Radiation Center, using the Rietan-FP program [21]. Magnetization data between 1.8 and 300 K were recorded using a Magnetic Properties Measurement System and MPMS-3 (both Quantum Design). Heat capacity measurements between 2 and 50 K were employed in the Physical Properties Measurement System (Quantum Design). Magnetization measurements up to 72 T were performed using a multilayered pulsed magnet with a duration of 4 ms. Typical pulsed magnetic field profiles are shown in Fig. 2. The magnetizations were measured at 1.4 K using the electromagnetic induction method employing a coaxial pick-up coil. Since it is difficult to obtain the absolute values of magnetization by this method, we calibrated the data to fit other magnetization curves measured on the same samples up to 7 T using an MPMS-3.

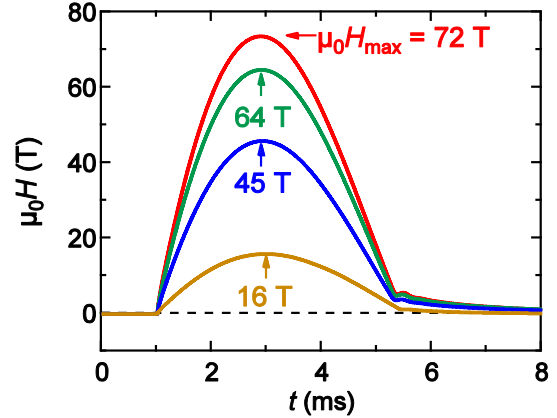

Fig. 2. Pulsed magnetic field profiles of magnetization measurements of $CuInCr_4S_8$ as a function of time.

## III. RESULTS AND DISCUSSION

### 3.1 Crystal Structure

Figure 3 shows the synchrotron powder XRD patterns of the $LiInCr_4S_8$, $LiGaCr_4S_8$, and $CuInCr_4S_8$ polycrystalline samples taken at room temperature and the results of their Rietveld analyses. We performed the Rietveld analyses by using a structural model with the cubic $F$–$43m$ space group as the main phase and $Cr_2S_3$ as a secondary phase in all cases, because there are some small peaks caused by a small amount of $Cr_2S_3$ impurity in all diffraction patterns, as seen in the inset of Fig. 3. The amount of $Cr_2S_3$ impurity in each sample is a few percent or less. In addition, the $LiGaCr_4S_8$ sample contains a minute amount of an unknown impurity phase. The lattice parameter obtained is $a$ = 10.13210(5), 9.96593(6), and 10.05970(11) Å for $LiInCr_4S_8$, $LiGaCr_4S_8$, and CuInCr$_4$S$_8$, respectively, which is consistent with that of a previous study [16]. There is no significant difference between the structural models with and without intersite defects between Li/Cu and Ga/In and vacancies. We estimated the amount of intersite defects by assuming some possible patterns, such as (Li$_{1-x}$In$_x$)(In$_{1-x}$Li$_x$)Cr$_4$S$_8$ for LiInCr$_4$S$_8$, yielding at most ~3% in all cases. Therefore, the crystallographic parameters obtained by assuming the occupancy of each site to be unity are listed in Table I. The differences between the Cr–Cr distances of the small and large tetrahedra are 9%, 7%, and 6% for LiInCr$_4$S$_8$, LiGaCr$_4$S$_8$, and CuInCr$_4$S$_8$, respectively, which are larger than those of LiInCr$_4$O$_8$ (5%) and LiGaCr$_4$O$_8$ (4%) [2], indicating stronger breathing in sulfides than in oxides.

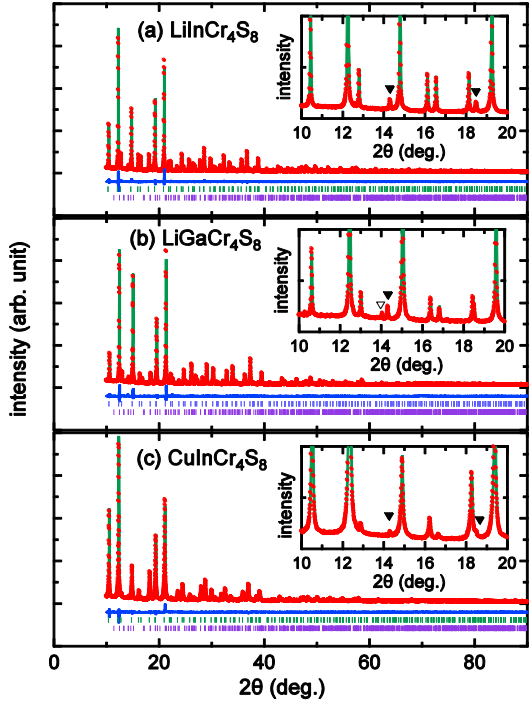

Fig. 3. Synchrotron powder XRD patterns of (a) LiInCr$_4$S$_8$, (b) LiGaCr$_4$S$_8$, and (c) CuInCr$_4$S$_8$ polycrystalline samples taken at room temperature. Filled circles are the experimental data. The overplotted curves show the calculated pattern, while the lower curve shows a difference plot between the experimental and calculated intensities. The upper and lower vertical bars indicate the positions of the Bragg reflections of the main and Cr$_2$S$_3$ impurity phases, respectively. The peaks in the inset indicated by the filled and open triangles are those of Cr$_2$S$_3$ and unknown impurities, respectively.

### 3.2 Magnetic Susceptibility at High Temperature

Figure 4(a) shows the temperature dependence of magnetization divided by the magnetic field $M/H$ of the LiInCr$_4$S$_8$, LiGaCr$_4$S$_8$, and CuInCr$_4$S$_8$ polycrystalline samples. The inverse of $M/H - \chi_{dia}$, where $\chi_{dia}$ is the diamagnetic contribution of core electrons ($\chi_{dia} = -9.2 \times 10^{-5}$, $-8.9 \times 10^{-5}$, and $-9.5 \times 10^{-5}$ cm$^{-3}$ mol-Cr$^{-1}$ for LiInCr$_4$S$_8$, LiGaCr$_4$S$_8$, and CuInCr$_4$S$_8$, respectively [23]), shown in Fig. 4(b), exhibits a linear temperature dependence above ~150 K for all samples, following the Curie-Weiss law $M/H = \chi = C/(T - \theta_W)$, where $\chi$, $C$, and $\theta_W$ are magnetic susceptibility, Curie constant, and Weiss temperature, respectively. $(M/H - \chi_{dia})^{-1}$ data of $LiInCr_4S_8$ and $LiGaCr_4S_8$ slightly deviate downward from the linear behavior below ~130 K owing to the presence of a trace amount of a ferrimagnetic impurity of $Cr_2S_3$ with a Curie temperature of 122 K [24]. $C$ and $\theta_W$ estimated by a Curie-Weiss fit to the $(M/H - \chi_{dia})^{-1}$ data between 200 and 300 K and the effective moment $p_{eff}$ and Lande g factor $g$ for $S = 3/2$ calculated from $C$ for the three sulfides are listed in Table II. The $g$ values close to 2 for all three sulfides indicate that the orbital moment is almost totally quenched and the three sulfides are $S = 3/2$ Heisenberg spin systems same as other Cr spinels.

Table 1. Crystallographic parameters for LiInCr$_4$S$_8$, LiGaCr$_4$S$_8$, and CuInCr$_4$S$_8$ determined by synchrotron powder XRD. The space group is $F\bar{4}3m$. The lattice parameter is $a$ = 10.13210(5), 9.96593(6), and 10.05970(11) Å, respectively. The thermal displacement parameter $B$ for Li in LiInCr$_4$S$_8$ and LiGaCr$_4$S$_8$ is constrained to 1.48, because inappropriate values were obtained in the refinement process probably due to the small atomic number of Li. See, for example, Ref. 22 for the definitions of the reliability factors $R_{wp}$, $R_p$, $R_e$, and $S$.

|  |  | x | y | z | g | B (Å) |
|---|---|---|---|---|---|---|
| LiInCr$_4$S$_8$ ($R_{wp}$ = 5.188, $R_p$ = 4.301, $R_e$ = 3.942, $S$ = 1.3160) | | | | | | |
| Li | 4a | 0 | 0 | 0 | 1 | 1.480 |
| In | 4d | 3/4 | 3/4 | 3/4 | 1 | 1.037(14) |
| Cr | 16e | 0.36966(8) x | x | x | 1 | 0.694(23) |
| S1 | 16e | 0.13524(10) x | x | x | 1 | 0.817(35) |
| S2 | 16e | 0.61093(11) x | x | x | 1 | 0.775(24) |
| LiGaCr$_4$S$_8$ ($R_{wp}$ = 4.208, $R_p$ = 3.382, $R_e$ = 3.052, $S$ = 1.3789) | | | | | | |
| Li | 4a | 0 | 0 | 0 | 1 | 1.480 |
| Ga | 4d | 3/4 | 3/4 | 3/4 | 1 | 0.923(23) |
| Cr | 16e | 0.37053(9) x | x | x | 1 | 0.823(23) |
| S1 | 16e | 0.13446(11) x | x | x | 1 | 0.732(35) |
| S2 | 16e | 0.61637(11) x | x | x | 1 | 0.884(28) |
| CuInCr$_4$S$_8$ ($R_{wp}$ = 3.985, $R_p$ = 3.281, $R_e$ = 3.115, $S$ = 1.2792) | | | | | | |
| Cu | 4a | 0 | 0 | 0 | 1 | 1.125(42) |
| In | 4d | 3/4 | 3/4 | 3/4 | 1 | 0.768(21) |
| Cr | 16e | 0.37120(15) x | x | x | 1 | 0.656(24) |
| S1 | 16e | 0.13402(17) x | x | x | 1 | 0.968(57) |
| S2 | 16e | 0.61194(13) x | x | x | 1 | 0.406(33) |

Table 2. Magnetic parameters of LiInCr$_4$S$_8$, LiGaCr$_4$S$_8$, and CuInCr$_4$S$_8$ obtained by a Curie-Weiss fit of $\chi$ between 200 and 300 K. The uncertainties in each column include those of the magnetization measurements and the Curie-Weiss fits.

|  | $C$ (cm$^3$ K mol-Cr$^{-1}$) | $p_{eff}$ ($\mu_B$) | $g$ | $\theta_W$ (K) |
|---|---|---|---|---|
| LiInCr$_4$S$_8$ | 1.70(6) | 3.69(7) | 1.91(3) | 3(1) × 10$^1$ |
| LiGaCr$_4$S$_8$ | 1.75(8) | 3.74(8) | 1.93(4) | −2(1) × 10$^1$ |
| CuInCr$_4$S$_8$ | 1.9(1) | 3.9(2) | 2.04(8) | −7(2) × 10$^1$ |

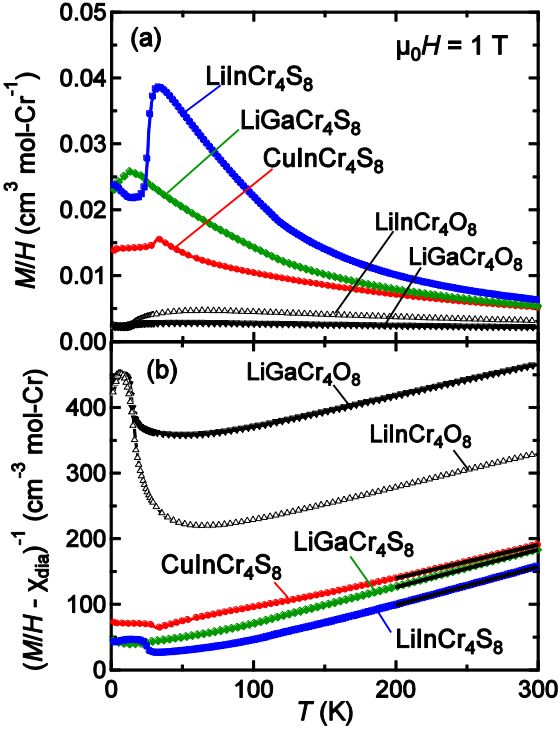

Fig. 4. (a) Temperature dependences of magnetization divided by magnetic field, $M/H$, measured in a magnetic field of $\mu_0 H = 1$ T for polycrystalline samples of $LiInCr_4S_8$, $LiGaCr_4S_8$, and $CuInCr_4S_8$. (b) Inverse of $M/H$ after subtraction of the diamagnetic contribution from core electrons, $\chi_{dia} = -9.2 \times 10^{-5}$, $-8.9 \times 10^{-5}$, and $-9.5 \times 10^{-5}$ cm$^3$ mol-Cr$^{-1}$ for $LiInCr_4S_8$, $LiGaCr_4S_8$, and $CuInCr_4S_8$, respectively [23]. The data of the polycrystalline samples of $LiInCr_4O_8$ and $LiGaCr_4O_8$ are also shown in (a) and (b) for comparison [2]. The solid lines in (b) show the results of a Curie-Weiss fit.

The Weiss temperatures are considerably different between the three sulfides. $\theta_W$ of $LiInCr_4S_8$ is positive, indicating that the ferromagnetic interaction is predominant, while those of $LiGaCr_4S_8$ and $CuInCr_4S_8$ are negative, indicative of the presence of predominant antiferromagnetic interaction. $\theta_W = -7(2) \times 10^1$ K for $CuInCr_4S_8$ is comparable to the $\theta_W$ values in the previous reports [17,18]. As will be discussed in the next section, antiferromagnetic $J$ and ferromagnetic $J'$ coexist in the breathing pyrochlore lattice of the three sulfides. It is expected that the summation of antiferromagnetic $J$ and ferromagnetic $J'$ results in the relatively small $|\theta_W|$, in contrast to the large $|\theta_W|$ of several hundred K in $LiInCr_4O_8$ and $LiGaCr_4O_8$ [2].

### 3.3 Relation between Structural Parameters and Magnetic Interactions

We will now discuss the relationship between the structural parameters and magnetic interactions in these three sulfides. Cr spinels are known to show a wide range of nearest-neighbor magnetic interactions from strongly antiferromagnetic to ferromagnetic, depending on the Cr–Cr distance [25]. This is because the nearest-neighbor magnetic interaction is determined by both the exchange interaction coming from the direct overlap of Cr 3$d$ orbitals and the superexchange interactions mediated by $p$ orbitals of anion X. The former is antiferromagnetic and rapidly increases with decreasing Cr–Cr distance. The latter is sensitive to the Cr–X–Cr angle but is largely insensitive to the Cr–Cr distance. In the Cr spinels, the latter interaction is expected to be ferromagnetic, because the Cr–X–Cr angles are close to 90°. The nearest-neighbor magnetic interactions in the Cr spinel oxides with shorter Cr–Cr distances are antiferromagnetic because the former is dominant, while those in the selenides and tellurides with larger Cr–Cr distances are ferromagnetic because the latter is dominant [25]. The sulfides are located between them and can be both antiferromagnetic and ferromagnetic. In $CdCr_2S_4$ and $HgCr_2S_4$, the ferromagnetic interaction is dominant, indicated by the large and positive $\theta_W$ exceeding 100 K [25]. Alternatively, $ZnCr_2S_4$, which has a smaller lattice constant than those of $CdCr_2S_4$ and $HgCr_2S_4$, has a much smaller $\theta_W$ of ~8 K, which becomes negative when the lattice constant becomes smaller by applying pressure [26–28]. However, we must take into account that the superexchange interaction via the Cr–S–S–Cr pathway, giving rise to the third neighbor interaction, can be moderate in Cr spinel sulfides. A theoretical study showed that this interaction is almost zero in Cr spinel oxides, but is several K in the sulfides [29].

Figure 5 shows the Weiss temperatures versus Cr–Cr distances of various Cr spinel sulfides, including $LiInCr_4S_8$, $LiGaCr_4S_8$, and $CuInCr_4S_8$. The nearest neighbor Cr–Cr distance $d_P = \sqrt{2}a/4$ for $ACr_2S_4$ (A = Zn, Cd, or Hg) [27] and the Cr–Cr distances $d_{BP}$ and $d'_{BP}$ on the small and large tetrahedra, shown in Fig. 1(b), respectively, and their average $\overline{d_{BP}} = (d_{BP} + d'_{BP})/2 = \sqrt{2}a/4$ for $LiInCr_4S_8$, $LiGaCr_4S_8$, and $CuInCr_4S_8$ are shown. As seen in Fig. 5, the differences between the $d_{BP}$ and $d'_{BP}$ of the three sulfides are considerably large, from 6% to 9%, resulting in significantly different $J$ and $J'$. The $d_{BP}$ values of these three sulfides are 3.40–3.45 Å, which are ~4% smaller than $d_{BP}$ = 3.53 Å for $ZnCr_2S_4$ [26]. We estimate the $J$ of the three sulfides by assuming that $\theta_W$ linearly increases with increasing Cr–Cr distances, yielding strongly antiferromagnetic $J$, corresponding to $\theta_W$ between −100 and −200 K [2,25]. In contrast, $d'_{BP}$ of the three sulfides is 3.65–3.74 Å, which is larger than $d_P$ = 3.63 Å of $HgCr_2S_4$, indicating that their $J'$ is strongly ferromagnetic, corresponding to $\theta_W$ = 150–300 K. Although it is expected in real materials that the third-neighbor interaction becomes several K, as discussed above [29], this result strongly suggests that the breathing pyrochlore magnets with alternating antiferromagnetic interaction $J$ and ferromagnetic interaction $J'$ could be realized with the three sulfides.

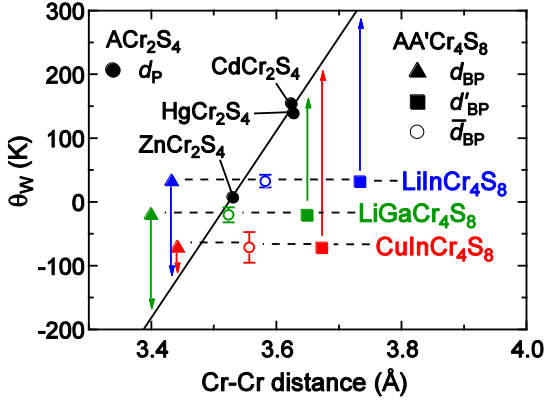

Fig. 5. Weiss temperature versus Cr–Cr distances for various Cr spinel sulfides. The values of $d_P$ for $ZnCr_2S_4$, $CdCr_2S_4$, and $HgCr_2S_4$ and $d_{BP}$, $d'_{BP}$, and $\bar{d}_{BP}$ = for $LiInCr_4S_8$, $LiGaCr_4S_8$, and $CuInCr_4S_8$ are indicated. The solid line is a linear fit to $\theta_W$ versus $d_P$ for $ZnCr_2S_4$, $CdCr_2S_4$, and $HgCr_2S_4$.

The values of $J$ and $J'$ are significantly different between the three sulfides. The $J'$ of $LiInCr_4S_8$ is expected to be much larger than $|J|$, because $d'_{BP}$ is considerably larger than those of the other two sulfides. In fact, $LiInCr_4S_8$ has a positive $\theta_W$, different from the other two sulfides. As seen in Fig. 5, the $|J|$ of $LiGaCr_4S_8$ seems to be comparable to $J'$, resulting in $\theta_W \sim 0$ owing to the cancelation of $J$ and $J'$. This result may correspond to the fact that the $\bar{d}_{BP}$ of $LiGaCr_4S_8$ is comparable to the $d_P$ of $ZnCr_2S_4$ ($\theta_W$ = 8 K). Alternatively, magnetic interactions in $CuInCr_4S_8$ show a different trend from those of other Cr spinel sulfides, including $LiInCr_4S_8$ and $LiGaCr_4S_8$. Although the $d_{BP}$ and $d'_{BP}$ of $CuInCr_4S_8$ are larger than those of $LiGaCr_4S_8$, respectively, $CuInCr_4S_8$ shows a negative $\theta_W$ of −70 K, which is considerably smaller than that of $LiGaCr_4S_8$, meaning that the magnetic interactions in $CuInCr_4S_8$ are more antiferromagnetic relative to its Cr–Cr distances than those in $LiGaCr_4S_8$. This difference between $CuInCr_4S_8$ and $LiGaCr_4S_8$ may be caused by the stronger antiferromagnetic superexchange interaction mediated by a Cu atom than that by a Li atom or weaker ferromagnetic superexchange interaction due to the larger Cr–S–Cr angle via the S2 site in $CuInCr_4S_8$ (97.7°) than that in $LiGaCr_4S_8$ (95.8°).

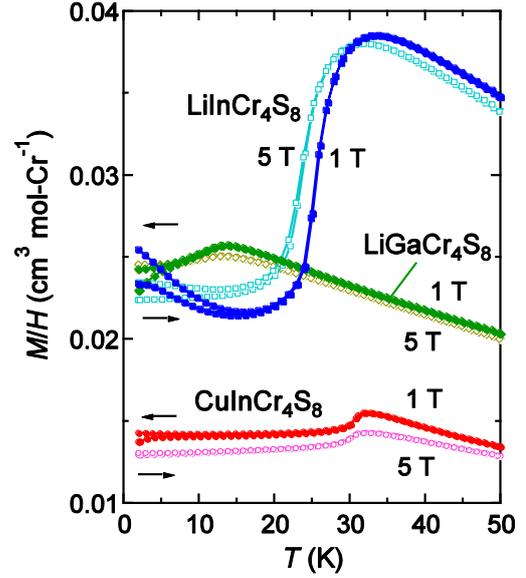

Fig. 6. Temperature dependences of field-cooled and zero-field-cooled magnetizations divided by the magnetic field $M/H$ of the $LiInCr_4S_8$, $LiGaCr_4S_8$, and $CuInCr_4S_8$ polycrystalline samples measured at magnetic fields of 1 (filled) and 5 T (open).

### 3.4 Magnetization and Heat Capacity at Low Temperature

At low temperatures, the magnetization $M$ and the heat capacity $C_p$ of the $LiInCr_4S_8$, $LiGaCr_4S_8$, and $CuInCr_4S_8$ polycrystalline samples exhibit a clear anomaly, as shown in Figs. 6 and 7, respectively. The $C_p/T$ of $LiInCr_4S_8$ exhibits a sharp peak at $T_p$ = 24 K, suggesting that a structural transition occurs at $T_p$, as in the case of Cr spinel oxides [2,4,30]. The strong decrease in $M/H$ at around $T_p$ suggests that the transition might be accompanied by an antiferromagnetic long-range order, the presence of which will be clarified by a future neutron diffraction or NMR experiment.

The anomalies in the $M/H$ and $C_p/T$ data of $LiGaCr_4S_8$ and $CuInCr_4S_8$ are smaller than that of $LiInCr_4S_8$, as seen in Figs. 6 and 7. The $M/H$ of $LiGaCr_4S_8$ measured in the magnetic field of 1 T shows a kink at 13 K. The $C_p/T$ of $LiGaCr_4S_8$ exhibits a broad peak at $T_p \sim 10$ K, as shown in Fig. 7(b), which is slightly lower than the peak temperature in $M/H$, suggesting that an antiferromagnetic long-range order occurs at around this temperature, even if there is a distribution of transition temperatures, which broadens the peak in $C_p/T$ data. The $M/H$ of CuInCr$_4$S$_8$ shows a sudden drop of ~10% at around 32 K with decreasing temperature, as is the case with the previous report, except that the temperature is slightly lower than that in the previous report of 35 K [20]. The $C_p/T$ of CuInCr$_4$S$_8$ shown in Fig. 7(b) exhibits a peak at $T_p$ = 28 K, which is also slightly lower than that in the previous report of 31 K, although the $C_p/T$ in the previous report shows two other peaks at 35 and 158 K [19]. These results confirm the presence of the previously reported antiferromagnetic long-range order, although the transition temperature in the present study is slightly lower than that in the previous report.

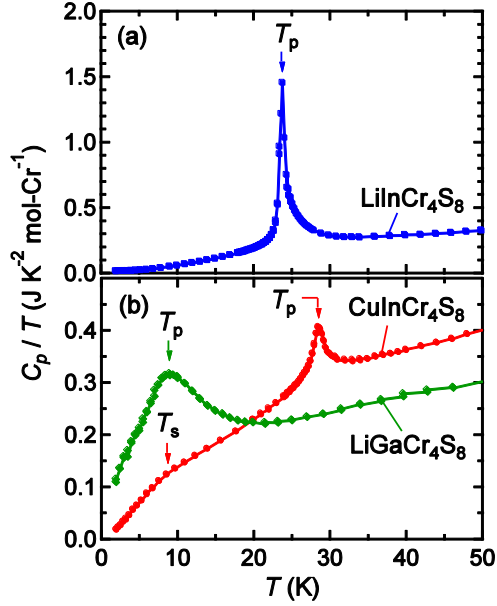

Fig. 7. Temperature dependence of heat capacity divided by temperature, $C_p/T$, of polycrystalline samples of (a) LiInCr$_4$S$_8$, (b) LiGaCr$_4$S$_8$, and CuInCr$_4$S$_8$.

### 3.5 High-Field Magnetization

In this section, we report high-field magnetization processes for the LiInCr$_4$S$_8$, LiGaCr$_4$S$_8$, and CuInCr$_4$S$_8$ powder samples measured up to 72 T using multilayered nondestructive pulsed magnets. We will discuss the magnetization curves ($M$–$H$ curves) of these three sulfides in the order of LiInCr$_4$S$_8$, LiGaCr$_4$S$_8$, and CuInCr$_4$S$_8$, and then compare them.

The $M$–$H$ curves of a LiInCr$_4$S$_8$ powder sample measured up to 41 and 56 T, shown in Fig. 8, have a large hysteresis loop closing at $\mu_0 H$ ~ 35 T. $M$ at 35 T is 2.6–2.7 $\mu_B$/Cr, exceeding 90% of the saturation magnetization of $M_s = gS\mu_B = 2.87$ $\mu_B$/Cr. $M$ becomes almost constant above 35 T, suggesting that all Cr$^{3+}$ spins are aligned in parallel above this magnetic field. This hysteresis loop does not seem to correspond to the formation of ferromagnetic domains, however, if we consider the high saturation field of 35 T and the metamagnetic-like behavior observed at $\mu_0 H$ ~ 30 and 20 T in the $M$–$H$ curves with increasing and decreasing magnetic fields, respectively. The small residual magnetization of 0.1–0.2 $\mu_B$/Cr for each curve might not be a spontaneous magnetization of a ferromagnet.

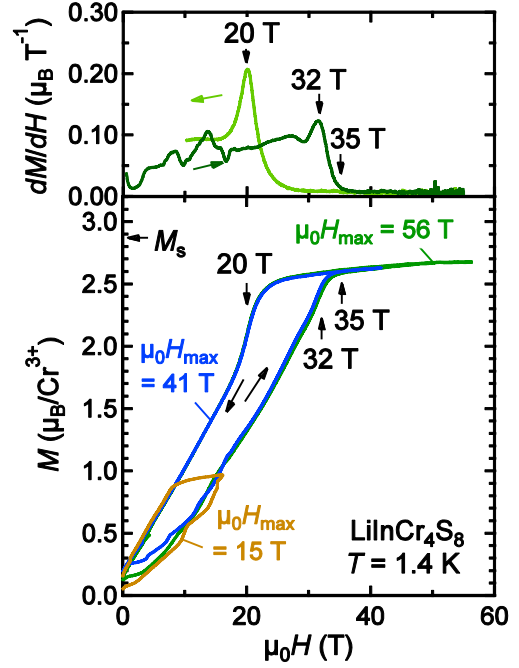

Fig. 8. Magnetization curves of a powder sample of LiInCr$_4$S$_8$. Measurements were performed up to 15, 41, and 56 T at 1.4 K using a multilayered pulsed magnet. The $dM/dH$ of the $M$–$H$ curve measured up to 56 T is shown at the top.

In contrast, the $M$–$H$ curves of LiGaCr$_4$S$_8$ do not show a hysteresis loop. Figure 9 shows the $M$–$H$ curves of a LiGaCr$_4$S$_8$ powder sample measured up to 72 T. The data measured up to 16, 45, 64, and 72 T completely overlap with each other and approach $M = 2.4$ $\mu_B$/Cr at $\mu_0 H = 72$ T, which is 85% of $M_s = 2.90$ $\mu_B$/Cr. A characteristic feature in the $M$–$H$ curves of LiGaCr$_4$S$_8$ is a gradual increase in $M$ at $\mu_0 H$ ~ 30 T. As seen in Fig. 9, the $M$ of LiGaCr$_4$S$_8$ almost linearly increases below 50 T. However, $dM/dH$ shows a minimum value at $\mu_0 H = 32$ T. $M$ at 32 T is 1.3 $\mu_B$/Cr, corresponding to 0.45 $M_s$, implying that this anomaly may be related to the formation of a half-magnetization plateau, although this $M$ is smaller than $M_s/2$ by approximately 10% and the magnetic field region corresponding to this anomaly is much narrower than those of the half-magnetization plateau in the Cr spinel oxides [8,31,32].

Figure 10 shows the $M$–$H$ curves and $dM/dH$ of a CuInCr$_4$S$_8$ powder sample measured up to 72 T. The $M$–$H$ curves show linear behavior below $\mu_0 H$ ~ 15 T, where the data measured up to 16, 45, 64, and 72 T completely overlap with each other, and then show a large hysteresis loop with a strong increase in $M$ above ~20 T with increasing $H$, which largely reproduces the previous result measured up to 38 T at 6 K [20]. The increase in $M$ is strongest at 25 T and becomes more gradual at around 40 T, which are clearly seen as a broad maximum and minimum in the $dM/dH$ data shown at the top of Fig. 10, respectively. $M$ strongly increases again at around 50 T, followed by a slowing down of the increase above 60 T. When $H$ decreases from 72 T, $M$ decreases with the similar $H$ dependence to that in the increasing process. The strongest decrease in $M$ appears at ~20 T and then $M$ coincides with that in the increasing process below ~15 T. In the previous study, the $M$–$H$ curves of CuInCr$_4$S$_8$ measured at 6 and 27 K up to the magnetic field of up to 38 T show a hysteresis loop above ~20 T [20], which is consistent with that in the present study. We find that this hysteresis is extended from a wide magnetic field region to a very high magnetic field exceeding 72 T.

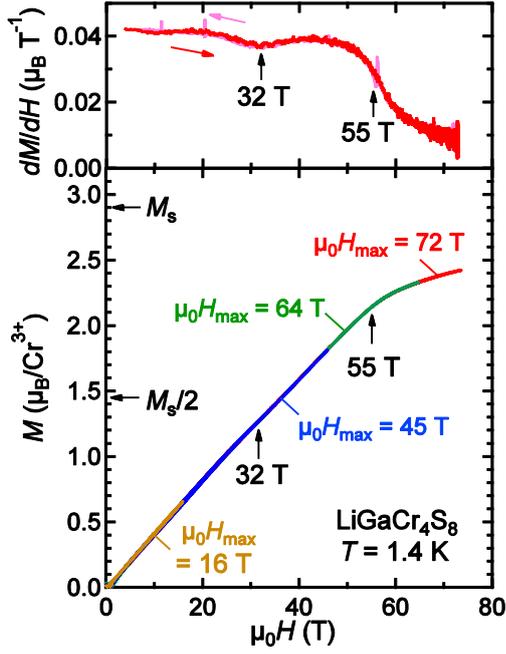

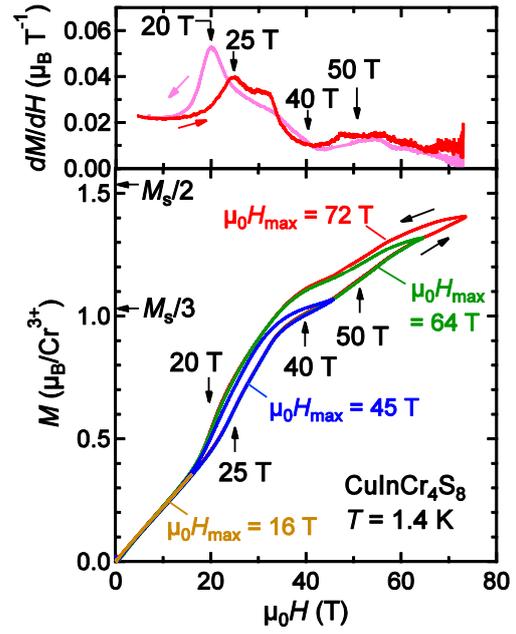

Fig. 9. Magnetization curves of a powder sample of LiGaCr$_4$S$_8$. Measurements were performed up to 16, 45, 64, and 72 T at 1.4 K using a multilayered pulsed magnet. The $dM/dH$ of the $M$–$H$ curve measured up to 72 T is shown at the top.

Fig. 10. Magnetization curves of a powder sample of CuInCr$_4$S$_8$. Measurements were performed up to 16, 45, 64, and 72 T at 1.4 K using a multilayered pulsed magnet. The $dM/dH$ of the $M$–$H$ curve measured up to 72 T is shown at the top.

The small change in $M$ at $\mu_0 H$ ~ 40 T in the $M$–$H$ curves of CuInCr$_4$S$_8$ is reminiscent of the formation of a magnetization plateau. The $M$ values at 40 T are 1.0 and 1.1 $\mu_B$/Cr in the increasing and decreasing processes of magnetic fields, respectively, both corresponding to approximately 1/3 of $M_s$, considering $g = 2.04$ and $S = 3/2$. In general, the $M$–$H$ curves of the antiferromagnets with triangular-based lattices, such as triangular or kagome lat-tices, often show a plateau at $M_s/3$, where the up-up-down spin arrangement for a triangle is stabilized in a finite magnetic field region [33−35]. Since such a 1/3 magnetization plateau has never been observed in pyrochlore magnets comprising tetrahedra, it would be interesting to know if this plateau is really formed in CuInCr$_4$S$_8$. Breathing pyrochlore antiferromagnets with $J'$ ~ 0 are expected to show stepwise $M$–$H$ curves at sufficiently low temperatures, as observed in Ba$_3$Yb$_2$Zn$_5$O$_{11}$ [11,36]. However, it is unlikely that CuInCr$_4$S$_8$ shows a stepwise $M$–$H$ curve by this scenario, because the $J'$ of CuInCr$_4$S$_8$ is expected to be ferromagnetic, as discussed in section 3.3, and this compound is suggested to show a magnetic long-range order at ~30 K, in contrast to Ba$_3$Yb$_2$Zn$_5$O$_{11}$, which shows spin-gap behavior down to the lowest measured temperature [10,11]. To clarify the formation mechanism of the plateau-like behavior at ~40 T in CuInCr$_4$S$_8$, it is necessary to obtain information on the magnetic structure by neutron scattering or NMR experiments conducted in magnetic fields. Alternatively, the gradual decrease in $dM/dH$ with increasing magnetic field above 60 T may be related to plateau formation at $M_s/2$. Magnetization measurements at higher magnetic fields will clarify whether the half magnetization plateau is formed in CuInCr$_4$S$_8$.

Figure 11 shows the magnetization curves of the three sulfides, where $M$ is normalized by $M_s = gS\mu_B$. That of $LiInCr_4O_8$ is also shown as a reference [8]. The slopes of the curves become steeper in the order of $\theta_W$, i.e., $LiInCr_4O_8$, $CuInCr_4S_8$, $LiGaCr_4S_8$, and $LiInCr_4S_8$. The $M$–$H$ curves of $LiInCr_4S_8$ and $CuInCr_4S_8$ show a large hysteresis loop with a metamagnetic increase in $M$, while there is no hysteresis in those of $LiGaCr_4S_8$ and $LiInCr_4O_8$, suggesting that the presence of the hysteresis is independent of the predominant magnetic interaction. In addition, the $M$–$H$ curve of $CuInCr_4S_8$ shows a plateau-like behavior at $M \sim M_s/3$, which does not appear in those of other three compounds, as shown in Fig. 11. We hope that the magnetic structure in the plateau-like region in $CuInCr_4S_8$ will be clarified by, for example, future neutron diffraction experiments in high magnetic fields.

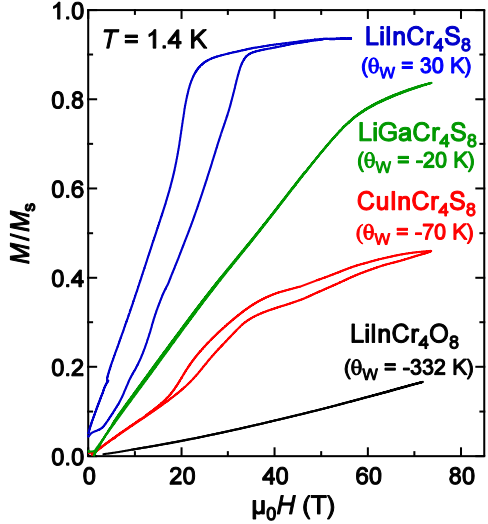

Fig. 11. Normalized magnetization curves of powder samples of $LiInCr_4S_8$, $LiGaCr_4S_8$, and $CuInCr_4S_8$ measured at 1.4 K. The data of a powder sample of $LiInCr_4O_8$ measured at 1.4 K are also shown for comparison [8].

### 3.6 Magnetic Ground States of Breathing Pyrochlore Sulfides

Finally, we will discuss the magnetic ground states of $LiInCr_4S_8$, $LiGaCr_4S_8$, and $CuInCr_4S_8$. As discussed in section 3.3, these three sulfides are suggested to be breathing pyrochlore magnets with antiferromagnetic $J$ and ferromagnetic $J'$. However, these three sulfides show considerably different magnetic properties at low temperatures. $LiInCr_4S_8$ shows a structural transition, which may be accompanied by a long-range magnetic order, indicated by a sharp peak in $C_p/T$ data and a strong decrease in $M/H$ at low temperatures. $LiGaCr_4S_8$ and $CuInCr_4S_8$ show smaller anomalies in $M$ and $C_p/T$ at low temperatures, as seen in Figs. 6 and 7, which might be due to the antiferro-magnetic long-range order. Moreover, the $M$–$H$ curves of the three sulfides measured up to 72 T by using pulsed magnetic fields have considerably different shapes, as seen in Fig. 11, although the slopes of the curves show the systematic variations with the predominant magnetic interactions.

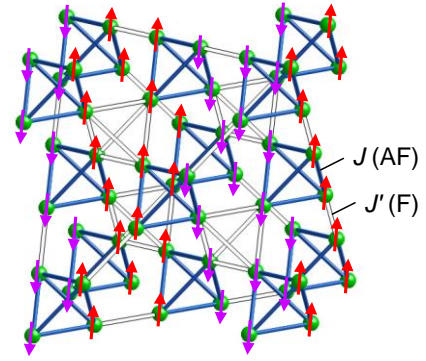

Fig. 12. A collinear spin structure comprising decoupled (100) ferromagnetic planes [15]. AF and F denote antiferromagnetic and ferromagnetic interactions, respectively.

In general, localized spin systems with both antiferromagnetic $J$ and ferromagnetic $J'$ often show a magnetic order where the spins coupled with $J$ and $J'$ are aligned in antiparallel and parallel configurations, respectively, such as in $NaNiO_2$ and $CaMnO_3$. However, such a magnetic order cannot be realized in the breathing pyrochlore case, because four spins in a large tetrahedron with $J'$ can be aligned in a parallel configuration but those in a small tetrahedron with $J$ cannot be aligned in an antiparallel configuration owing to the geometrical frustration. A previous theoretical study indicated that a collinear order comprising decoupled (100) ferromagnetic planes, shown in Fig. 12 [15], where spins on a $J'$ tetrahedron are aligned in parallel, while antiparallel and parallel pairs coexist on a $J'$ tetrahedron, is a ground state in the classical Heisenberg spin case. The spin structure model of $CuInCr_4S_8$ proposed in the previous neutron diffraction study, where magnetic Bragg peaks indicating the presence of a long-range magnetic order are observed at 4.2 K, is consistent with this magnetic structure [19]. Different magnetic properties at low temperatures between the three sulfides suggest that the magnetic ground states of $LiInCr_4S_8$ and $LiGaCr_4S_8$ are different from it. It is a remarkable point for the breathing pyrochlore sulfides that the different atoms at the tetrahedral site might give rise to a wide variety of magnetic ground states, although they commonly have antiferromagnetic $J$ and ferromagnetic $J'$. There are also unsolved issues in the magnetic properties of the three sulfides, such as the formation mechanism of the plateau-like behavior at around $M_s/3$ in the $M$–$H$ curves of $CuInCr_4S_8$. We hope that they will be clarified by various methods, such as future neutron scattering and NMR measurements.

## IV. CONCLUSIONS

The structural and magnetic properties of three breathing pyrochlore magnets, LiInCr$_4$S$_8$, LiGaCr$_4$S$_8$, and CuInCr$_4$S$_8$, are investigated. The structural parameters determined by the powder synchrotron XRD data and the temperature dependence of magnetization strongly suggest that these three sulfides have a unique breathing pyrochlore lattice with alternating antiferromagnetic $J$ and ferromagnetic $J'$. The magnetic properties are more antiferromagnetic in the order of CuInCr$_4$S$_8$, LiGaCr$_4$S$_8$, and LiInCr$_4$S$_8$. LiInCr$_4$S$_8$ shows a sharp peak at 24 K in the $C_p/T$ data, accompanied by a strong decrease in $M/H$, indicating that a structural transition occurs at this temperature. The $M-H$ curves of LiInCr$_4$S$_8$ show a large hysteresis loop. LiGaCr$_4$S$_8$ has a small $|\theta_W|$ due to the cancellation of antiferromagnetic $J$ and ferromagnetic $J'$. The $M-H$ curves of LiGaCr$_4$S$_8$ do not show a hysteresis loop that is different from other two sulfides, but show a small anomaly at around $M = M_s/2$, which may be related to the half magnetization plateau. The $M-H$ curves of CuInCr$_4$S$_8$, which is more antiferromagnetic than the other two sulfides, show metamagnetic behavior above ~20 T with a large hysteresis loop, where a plateau-like behavior appears at around $M_s/3$. Thus, these three sulfides show significantly different magnetic properties, although they commonly have both antiferromagnetic $J$ and ferromagnetic $J'$ on their pyrochlore lattices, indicating that rich physics exists in antiferromagnetic-ferromagnetic breathing pyrochlore magnets.

## ACKNOWLEDGMENTS

We thank T. Yajima for helpful discussions. This work was partly carried out at the International MegaGauss Science Laboratory and Materials Design and Characterization Laboratory under the Visiting Researcher Program of the Institute for Solid State Physics, the University of Tokyo and supported by JSPS KAKENHI Grant Number 16H03848. The synchrotron X-ray diffraction experiments were conducted at BL5S2 of the Aichi Synchrotron Radiation Center, Aichi Science and Technology Foundation, Aichi, Japan (Proposal Nos. 201703028 and 201702049).

*Electronic mail: yokamoto@nuap.nagoya-u.ac.jp